\documentclass[prb,preprintnumbers,amsmath,amssymb,twocolumn]{revtex4}

\usepackage{graphicx}
\usepackage{dcolumn}
\usepackage{bm}
\usepackage{latexsym,epsfig}
\usepackage{chemarrow}

\begin{document}

\title{Coupled First-Order Transitions In A Fermi-Bose Mixture}

\author{K Sheshadri$^{1,}$} \email{kshesh@gmail.com}  
\author{A Chainani$^{2,}$} \email{chainania@gmail.com}
\affiliation{ $^{1}$226, Bagalur, Bangalore North, Karnataka State, India 562149 }

\affiliation{ $^{2}$Condensed Matter Physics Group, National Synchrotron Radiation Research Center, Hsinchu 30076, Taiwan }

\date{\today}

\begin{abstract}

A model of a mixture of spinless fermions and spin-zero hardcore bosons, with filling fractions $\rho_F$ and $\rho_B$, respectively, on a two-dimensional square lattice with {\em composite} hopping $t$ is presented. In this model, hopping swaps the locations of a fermion and a boson at nearest-neighbor sites. When $\rho_F+\rho_B=1$, the fermion hopping amplitude $\phi$ and boson superfluid amplitude $\psi$ are calculated in the ground state within a mean-field approximation. The Fermi sector is insulating ($\phi=0$) and the Bose sector is normal ($\psi=0$) for $0 \le \rho_F < \rho_c$. The model has {\em  coupled first-order} transitions at $\rho_F = \rho_c \simeq 0.3$ where both $\phi$ and $\psi$ are discontinuous. The Fermi sector is metallic ($\phi>0$) and the Bose sector is superfluid ($\psi>0$) for $\rho_c < \rho_F < 1$. At $\rho_F=1/2$, fermion density of states $\rho$ has a van Hove singularity, the bulk modulus $\kappa$ displays a cusp-like singularity, the system has a density wave (DW) order, and $\phi$ and $\psi$ are maximum. At $\rho_F=\rho_{\kappa} \simeq 0.81$, $\kappa$ vanishes, becoming {\em negative} for $\rho_{\kappa}<\rho_F<1$. The role of composite hopping in the evolution of Fermi band dispersions and Fermi surfaces as a function of $\rho_F$ is highlighted. The estimate for BEC critical temperature is in the subkelvin range for ultracold atom systems and several hundred kelvins for possible solid-state examples of the model.

\end{abstract}



\maketitle

\section{Introduction}
\label{intro}

Fermi-Bose mixtures (FBMs) constitute an unusual phase of matter, the earliest examples of which are the mixed phase of type-II superconductors and He$^3$-He$^4$ mixtures \cite{Ebner}. In the past two decades, fascinating experiments on FBMs of ultracold atoms have revealed unique properties and given a major thrust to their study \cite{Truscott, Schreck, Hadzibabic}. While superfluidity was observed very early in either the Fermi or Bose sector of an FBM, the coexistence of superfluidity in both sectors was reported very recently \cite{Ferrier}.

Theoretical and experimental studies of FBMs have developed significantly over the years, particularly in terms of a crossover between the limiting cases of the BCS picture of superconductivity and the BEC picture of superfluidity \cite{Randeria, Ketterle}. The BCS-BEC crossover was predicted to occur for excitons in semiconductors \cite{Keldysh}, quarks \cite{Kerbikov}, and interestingly, it was first realized in ultracold fermionic atoms with s-wave interactions \cite{Bartenstein}. The growing number of experimental results includes the Feshbach resonance across the BCS-BEC crossover \cite{Regal}, formation of a Feshbach molecule in an FBM \cite{Cumby}, and the role of three body physics in describing an FBM \cite{Bloom} to name a few.  More recently, experimental results of some iron-based superconductors and the relevance of their electronic structure and properties \cite{Lubashevsky, Okazaki, Kasahara, Rinott} have also been discussed in relation to theoretical results of the BCS-BEC crossover \cite{Randeria, Randeria2, quick}. While it is well-known that interactions between fermions mediated by phonons are at the root of the BCS theory of superconductivity, several studies have also considered their importance in FBMs of ultracold atoms \cite{Bijlsma, Viverit, Capuzzi, Albus}. The BCS theory was preceded by the so-called Boson-Fermion (BF) model \cite{Schafroth}, which discusses itinerant fermions interacting with localized bosons composed of bound pairs of fermions of opposite spins. The BF model was later applied to study electrons interacting with local lattice deformations \cite{Ranninger} and high temperature superconductivity \cite{Ranninger2, Friedberg, Gershkenbein, Domanski}. More recently, it was also used for describing resonance superfluids in the BCS-BEC crossover regime \cite{Shin} as well as a temperature driven BCS-BEC crossover in an FBM \cite{Maska2}. The interplay of bosonic and fermionic degrees of freedom is therefore of great importance in several physical systems.


In this work, we consider a model of spinless fermions and hardcore bosons on a square lattice with a filling constraint of one particle per site. 
While earlier studies on FBMs, for example by Kuklov and B. Svistunov\cite{Kuklov}, have considered the same constraint of one particle per site, they considered only fermions and bosons hopping independently, and included on-site Coulomb interaction for fermions, bosons and between fermions and bosons.  Very interestingly, the authors could show that the FBM in a commensurate optical lattice showed counterflow superfluidity. Further, while the dynamics of FBMs in the Mott phase has also been discussed in the literature, long range density wave phases have been identified for fermion-fermion and boson-boson hopping in the presence of on-site boson-boson and boson-fermion Coulomb interaction \cite{Lewenstein, Pollet}. However,
to our knowledge, a  {\em composite}  hopping that involves an exchange of a fermion with a boson, when they occupy neighboring sites, has not been reported to date. This form of hopping distinguishes our work from earlier work on FBMs. We allow no independent hopping. Indeed, such a hopping would be forbidden in our model as it would result in either two spinless fermions or two hardcore bosons at a site. A seemingly simple realization would be a model of strongly correlated electrons on a lattice, with an excess of electrons over lattice sites (for example, one-band Hubbard model above half filling), resulting in both singly-occupied and doubly-occupied sites. A pair of fermions on a site could be treated as a boson \cite{Maska2}. When an electron hops from a doubly-occupied site to a singly-occupied site, a fermion and a boson are swapped. The composite hopping mechanism provides a good description of this situation. In addition, our work is able to address the case of coupled transitions in an FBM due to the composite hopping, as we show in the following. 

We briefly summarize our main results. We perform a mean-field theory of our model at zero temperature. The model displays two distinct phases separated by {\em coupled first-order} transitions at Fermi filling fraction $\rho_F=\rho_c\simeq 0.3$: for $\rho_F<\rho_c$ the Fermi sector is insulating and the Bose sector is a normal gas, while for $\rho_F>\rho_c$ the Fermi sector is metallic and the Bose sector is a superfluid. The fermion band width varies with the Fermi energy. The fermion density of states and bulk modulus exhibit cusp-like singularities at $\rho_F = 1/2$, where the Fermi and Bose sectors have a density-wave (DW) ordering coexisting with maximum metallicity and superfluidity. The bulk modulus becomes {\em negative} for $\rho_F>\rho_{\kappa} \simeq 0.81$. Our estimate of BEC transition temperature in this model is a fraction of the composite hopping strength, and could be as high as several hundred kelvins in solids while it may be expected to be in the subkelvin range in ultracold atom systems.

\section{Hamiltonian}
\label{ham}

The Hamiltonian we study is
\begin{eqnarray}
\label{eq:ham}
H &=& -\alpha \sum_{i} \left[b_i^{\dagger}b_i + f_i^{\dagger}f_i - 1\right]  -\mu \sum_{i} f_i^{\dagger}f_i   \nonumber \\
&& + \frac{U}{2} \sum_{i} b_i^{\dagger}b_i (b_i^{\dagger}b_i - 1) - t \sum_{<ij>} f_i^{\dagger}f_j b_j^{\dagger} b_i
\end{eqnarray}
where $b_i^{\dagger}$ creates a spin-zero boson and $f_i^{\dagger}$ creates a spinless fermion at site $i$ of a two-dimensional square lattice. The parameter $\alpha$ is a Lagrange multiplier: it is used for imposing the condition that the total number of fermions and bosons be a certain given number (taken to be equal to the number of lattice sites in this work); $\mu$ is the chemical potential used to determine $\rho_F$; $U$ is the repulsive energy between two bosons at the same site; finally, $t$ is the strengh of the {\em composite} hopping: in this process a fermion hops from a site $j$ to a neighboring site $i$ when a boson located there simultaneously hops to site $j$. The boson operators satisfy $[ b_i, ~b_j ] = 0=[ b_i^{\dagger}, ~b_j^{\dagger} ], ~[ b_i, ~b_j^{\dagger} ]  = \delta_{ij}$, while the spinless-fermion operators satisy
$\{ f_i, ~f_j \} = 0, ~\{ f_i^{\dagger}, ~f_j^{\dagger} \}=0, ~\{ f_i, ~f_j^{\dagger} \}  = \delta_{ij}$, where $\left[ a, b \right] = ab-ba$ and $\{ a, b \} = ab+ba$. In this paper, we restrict ourselves to hardcore bosons, i.e. the limit $U \to \infty$, and impose the constraint
\begin{equation}
\label{eq:fill1}
\rho_F + \rho_B = 1,
\end{equation}
($\rho_B$ is the boson filling fraction) by requiring $\partial F/\partial \alpha = 0$, $F$ being the free energy. Given the nature of the composite hopping term, a site that is unoccupied by a fermion or a boson remains unoccupied, and a site that is doubly occupied by a fermion and a boson remains doubly occupied. 
The only interesting case therefore is one where the constraint (\ref{eq:fill1}) operates, with each site being occupied either by a fermion or a boson.

\section{Calculation}
\label{calc}

To calculate the properties of the hamiltonian (\ref{eq:ham}) , we use the following approximation: 

\begin{enumerate}
	\item If $\hat{F}$ is a fermion operator and $\hat{B}$ is a boson operator, then $\hat{F}\hat{B} \simeq\langle \hat{F} \rangle \hat{B} + \langle \hat{B} \rangle \hat{F} - \langle \hat{F} \rangle \langle \hat{B} \rangle$. 

	\item If $\hat{B}_i, ~ \hat{B}_j$ are boson operators at sites $i \ne  j$, then $\hat{B}_i \hat{B}_j \simeq \langle \hat{B}_i \rangle \hat{B}_j + \langle \hat{B}_j \rangle\hat{B}_i - \langle \hat{B}_i \rangle \langle \hat{B}_j \rangle$.

\end{enumerate}
The first step implies $\langle \hat{F}\hat{B} \rangle \simeq \langle \hat{F} \rangle \langle \hat{B} \rangle$, so this is equivalent to factorizing the state space into a product of fermion and boson subspaces. The second step implies $\langle \hat{B}_i \hat{B}_j \rangle \simeq \langle \hat{B}_i \rangle \langle \hat{B}_j \rangle$, which is equivalent to further factorizing the boson subspace into a product of single-site subspaces: this is the standard mean-field decoupling used in the bosonic Hubbard model \cite{Sheshadri, PRL95, FisherWeichman89, gutz, spin1_1, spin1_2, spin1_3}. When we follow these steps, the composite hopping term is transformed according to 
\begin{eqnarray}
\label{eq:mfa}
f_i^{\dagger}f_j b_j^{\dagger} b_i &\simeq& \langle f_i^{\dagger}f_j \rangle ~(\langle b_j^{\dagger} \rangle b_i + \langle b_i \rangle b_j^{\dagger} - \langle b_j^{\dagger} \rangle \langle b_i \rangle)  \nonumber \\
&& + \langle b_j^{\dagger} \rangle \langle b_i \rangle f_i^{\dagger}f_j - \langle f_i^{\dagger}f_j \rangle \langle b_j^{\dagger} \rangle \langle b_i \rangle,
\end{eqnarray}
and the hamiltonian is approximated by
\begin{eqnarray}
\label{eq:mfham}
H^{MF} &=& H_0 + H_1 + H_2,  ~~\mathrm{where} \nonumber \\
H_0 &=& N (2 \phi \psi^2 + \alpha),  \nonumber \\
H_1 &=& - (\alpha+\mu)\sum_{i} f_i^{\dagger}f_i - \frac{1}{z}\psi^2 \sum_{<ij>} f_i^{\dagger}f_j, ~ \mathrm{and}  \nonumber \\
H_2 &=& \sum_{i} \left[ -\alpha b_i^{\dagger}b_i - \phi \psi (b_i+b_i^{\dagger}) \right].
\end{eqnarray}
We have taken $zt=1$ ($z$ is the coordination number of the lattice), and introduced the ground-state expectation values
\begin{equation}
\label{eq:sce}
\phi = \langle f_i^{\dagger} f_j \rangle, ~~ \psi = \langle b_i \rangle = \langle b_j^{\dagger} \rangle.
\end{equation}
We assume $\phi ~\mathrm{and} ~\psi$ to be real and homogeneous. We consider $\psi$ to be the boson superfluid amplitude \cite{FisherWeichman89, Sheshadri, PRL95} and $\phi$ to be fermion hopping amplitude. It can be observed from the expression for $H_1$ in equation (\ref{eq:mfham}) that when $\psi=0$, the fermion hopping term vanishes, so $\phi=0$. This indicates that when there is a superfluid transition in the Bose sector, there is an accompanying metal-insulator transition in the Fermi sector, resulting in the two transitions being always coupled. We have dropped the $U$-term: the hardcore boson limit $U \to \infty$ is incorporated by taking the single-site boson occupation number basis $\{ |0\rangle, ~ |1\rangle \}$ for diagonalizing the single-site $2 \times 2$ matrix
\begin{equation}
\label{eq:matrix}
h_2 = 
\begin{bmatrix}
    0       & -\phi \psi  \\
    -\phi \psi       & -\alpha 
\end{bmatrix}
\end{equation}
 of $H_2/N$, with eigenvalues
\begin{equation}
\label{eq:bspect}
\lambda_{\pm} = \frac{1}{2} \left[ -\alpha \pm R \right], ~~\mathrm{where} ~~R  =\sqrt{\alpha^2 + 4\phi^2\psi^2}.
\end{equation}
To solve the fermion sector hamiltonian $H_1$, we move over to $k$-space using the Fourier transform $f_i = N^{-1/2} \sum_{\bf k} e^{i {\bf k.r_i}} f_{\bf k}$, so that
\begin{equation}
\label{eq:h1ft}
H_1 = \sum_{\bf k} (\varepsilon_{\bf k}-\mu) f_{\bf k}^{\dagger} f_{\bf k},
\end{equation}
where
\begin{equation}
\label{eq:dispersion}
\varepsilon_{\bf k} = -\alpha - \psi^2 \gamma_{\bf k} ~\mathrm{and} 
~\gamma_{\bf k} = 2 (\cos k_x + \cos k_y)/z.
\end{equation}
The zero-temperature free energy per lattice site $F = H_0/N + \langle H_1 \rangle/N + \lambda_{-}$ is now
\begin{equation}
\label{eq:fe}
F  = 2 \phi \psi^2 + \frac{1}{2} (\alpha - R) + \frac{1}{N}\sum_{\bf k} (\varepsilon_{\bf k}-\mu) \langle f_{\bf k}^{\dagger} f_{\bf k} \rangle.
\end{equation}
We observe that $(1/N)\sum_{\bf k} \langle f_{\bf k}^{\dagger} f_{\bf k} \rangle = \rho_F$. Using the definition of $\phi$ in (\ref{eq:sce}) and going over to $k$-space, we get 
\begin{equation}
\label{eq:phi1a}
\phi = \frac{1}{Nz} \sum_{<ij>} \langle f_i^{\dagger} f_j \rangle = \frac{1}{N} \sum_{\bf k} \gamma_{\bf k} \langle f_{\bf k}^{\dagger} f_{\bf k} \rangle,
\end{equation}
so that
\begin{equation}
\label{eq:fe2}
F  = \frac{1}{2} (\alpha - R) + \phi \psi^2 - (\mu + \alpha) \rho_F.
\end{equation}
Introducing the Fermi density of states
\begin{equation}
\label{eq:dos}
\rho (E) = \frac{1}{N} \sum_{\bf k} \delta(E-\varepsilon_{\bf k}),
\end{equation}
we can write
\begin{equation}
\label{eq:phi3}
\phi = -\frac{1}{\psi^2}\int_{E_0}^{\mu} dE ~(\alpha + E) \rho(E),
\end{equation}
where $\mu$ is chosen such that the fermion filling fraction
\begin{equation}
\label{eq:rhof}
\rho_F = \int_{E_0}^{\mu} dE \rho(E)
\end{equation}
has a desired value. Here, $E_0 = -\alpha-\psi^2$ is the minimum value of fermion energy. To calculate the density of states (\ref{eq:dos}),  we convert the $k$-sum into an integral according to ~$(1/N)\sum_{\bf k} \to (1/4\pi^2) \int d{\bf k}$. Since $\varepsilon_{\bf -k} = \varepsilon_{\bf k}$, the $k$-space integral is four times the integral over the first quadrant of the Brillouin zone, and so we have
\begin{equation}
\label{eq:dos2}
\rho (E) = \frac{1}{\pi^2} \int_0^{\pi} dk_x \int_0^{\pi} dk_y ~\delta (E+\alpha+\psi^2\gamma_{\bf k}).
\end{equation}
The integral over $k_y$ can be easily evaluated, and we get
\begin{eqnarray}
\label{eq:dos3}
\rho (E) &=& \frac{2}{\pi^2\psi^2} f\left( \frac{\alpha + E}{\psi^2} \right), ~\mathrm{where} \nonumber \\
f(u) &=& \int_0^{\pi} \frac{dk_x}{\sqrt{1-(2 u + \cos k_x)^2}}.
\end{eqnarray}

We can readily see that the function $f(u)$ is real only when $-1 \le u \le 1$, and is non-negative. Therefore we have the inequality $-\alpha-\psi^2 \le E \le -\alpha+\psi^2$ for the fermion energy $E$. The Fermi band width is therefore $2\psi^2$. We can also see that $f(0) \rightarrow \infty$, and this is the van Hove singularity at $\rho_F=1/2$. We substitute the above expression for $\rho (E)$ in to equations (\ref{eq:phi3}) and (\ref{eq:rhof}) and transform the integrals to obtain
\begin{equation}
\label{eq:rhofphi}
\rho_F = \frac{2}{\pi^2}\int_{-1}^{u_F} f(u)  du, ~~\phi = -\frac{2}{\pi^2}\int_{-1}^{u_F} u f(u)  du,
\end{equation}
where $u_F=(\alpha+\mu)/\psi^2$. For a given $(\mu, T)$, we determine $\psi$ and $\alpha$ by simultaneously solving $\partial F/\partial \psi = 2 \psi \phi ( 1-\phi/R ) = 0$ and $\partial F/\partial \alpha = (1/2) [ (1-2\rho_F) - \alpha/R  ] = 0$. We obtain two solutions for $\psi$, namely $\psi = 0$ (disordered, or normal Bose phase) and $R = \phi$ (ordered, or superfluid phase), and the solution $\alpha = (1-2\rho_F) R$. We can write this as $\alpha^2 [1 - (1-2\rho_F)^2] = 4\phi^2\psi^2(1-2\rho_F)^2$, so $\alpha=0, ~ R=0$, and the minimum free energy $F_0=0$ in the disordered phase; in the ordered phase, $\alpha = (1-2\rho_F)\phi, ~\psi^2 =\rho_F (1-\rho_F), ~\mathrm{and}~ F_0=\Delta$. Here $\Delta=\rho_F^2[-\phi + u_F(\rho_F-1)]$ is the minimum free energy of the ordered phase. Since $F_0=0$ in the disordered phase, we can interpret $\Delta$ as energy cost of creating superfluidity. Since $u_F<0$ for $\rho_F<0.5$, it is clear that there is a certain $\rho_c < 0.5$ such that $\Delta > 0$ for $\rho_F < \rho_c$.

\section{Discussion}
\label{discuss}

\begin{figure}
\includegraphics[width=0.45 \textwidth]{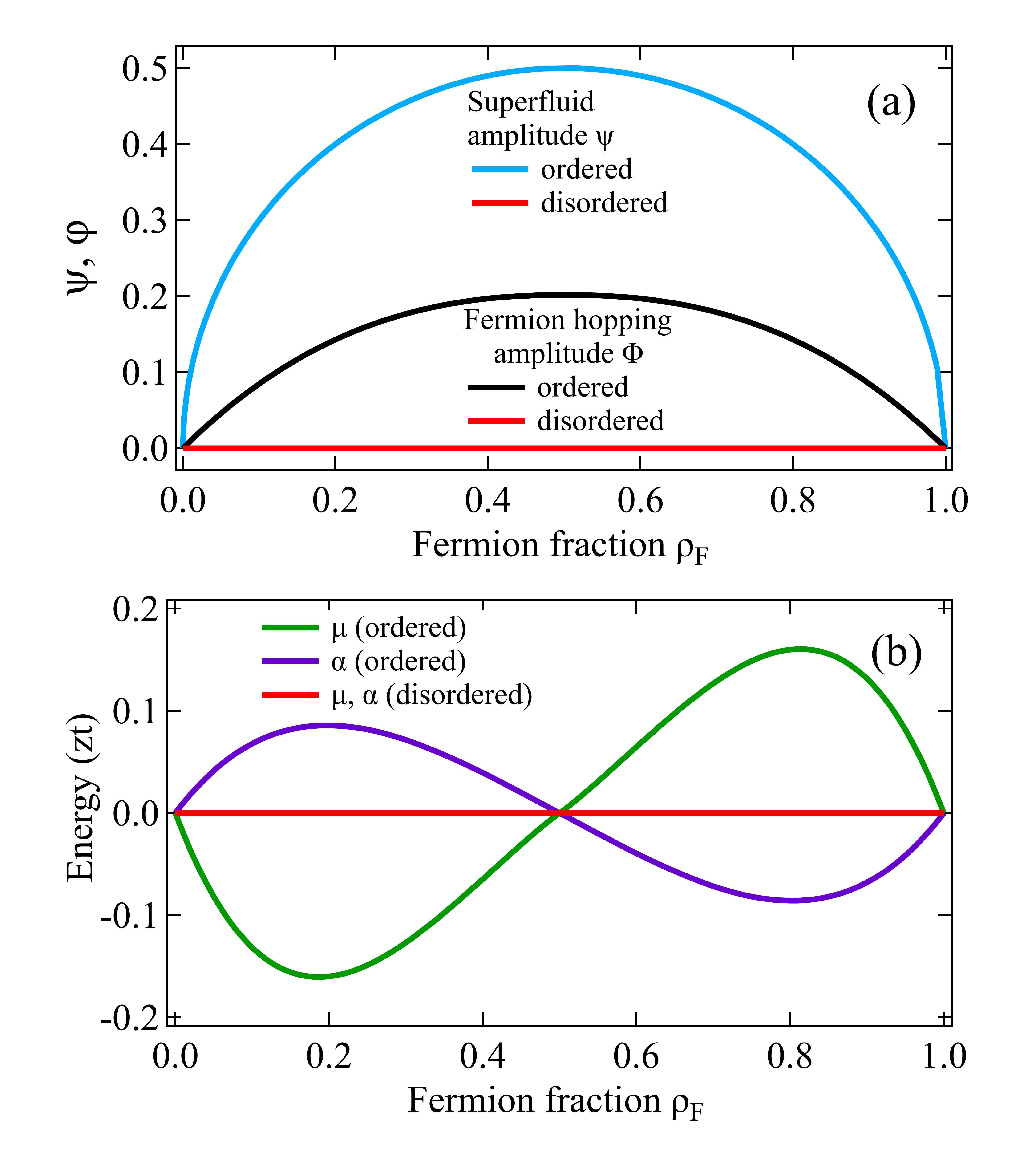}
\caption{(a) Plot of the order parameter $\psi$ (blue) and $\phi$ (black) as a function of $\rho_F$ for the ordered solution. The red line shows both in the disordered phase. (b) Plot of the chemical potential $\mu$ (green) and  $\alpha$ (blue) in the ordered phase. The red line shows both in the disordered phase. The non-monotonic behavior of $\alpha$ in the ordered phase is determined by the filling constraint (Eq.(\ref{eq:fill1})), which is also responsible for a similar non-monotonic behavior of $\mu$ in the ordered phase.}
\label{fig:fig1}
\end{figure}

\begin{figure}
\includegraphics[width=0.45 \textwidth]{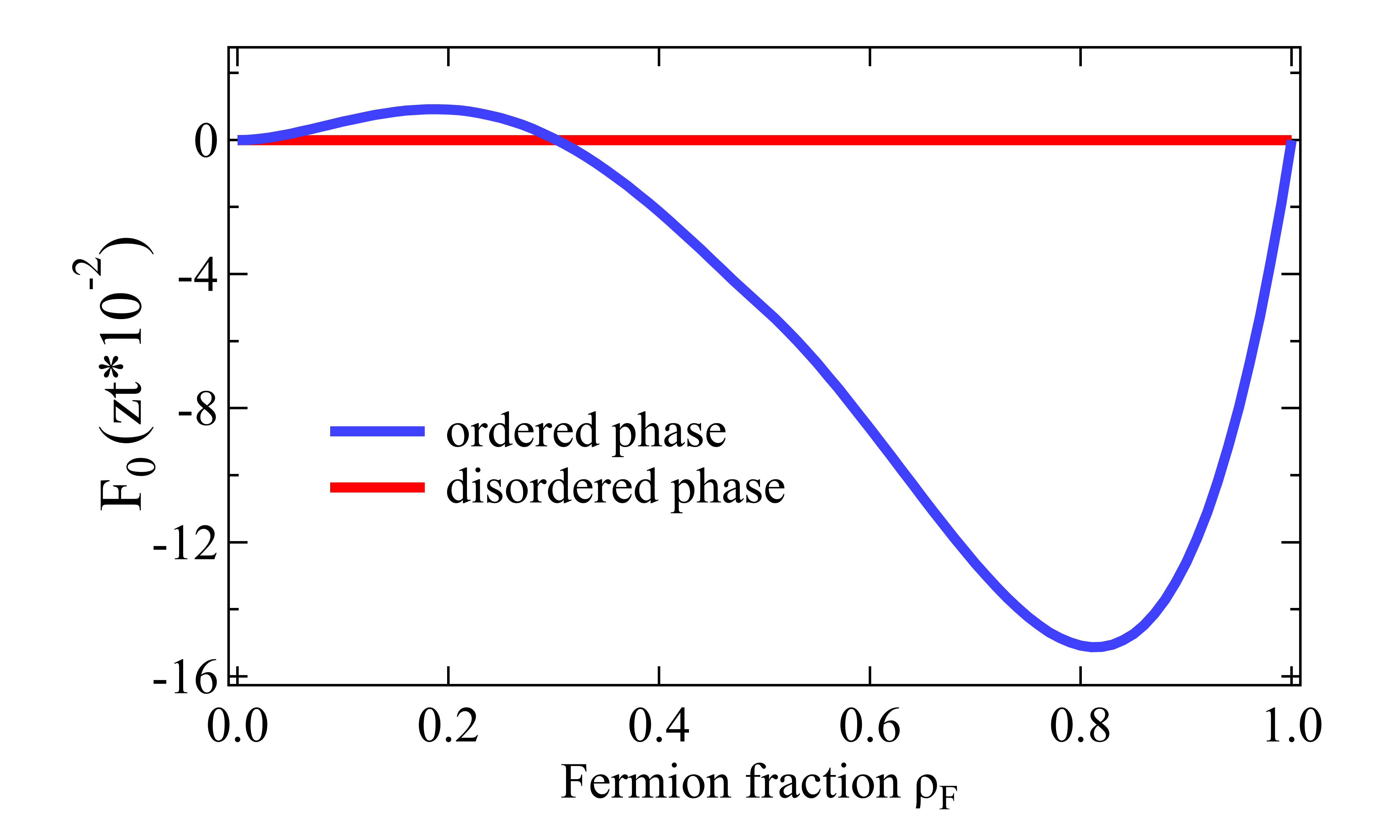}
\caption{Plot of the minimum free energy $F_0$ for the ordered (blue) and disordered (red) phases. In the ordered phase, the free energy is the cost $\Delta$ of superfluid state, and becomes positive for $0<\rho_F<0.3$, favoring the insulating normal Bose phase. At $\rho_F=0.3$ the free energy thus has a derivative discontinuity and we have a coupled first-order transition.}
\label{fig:fig2}
\end{figure}

As mentioned above, the self-consistency equation for the order parameter $\psi$ has two solutions. For each of these solutions, we compute the quantities $\phi, \alpha$ and $\mu$. Fig.\ref{fig:fig1} shows the plots of $\psi, \phi$ (panel (a)), $\alpha$ and $\mu$ (panel (b)) as functions of $\rho_F$.

The correct solution at each filling $\rho_F$ is determined based on the minimum free energy $F_0$. Fig.\ref{fig:fig2} shows plots of $F_0$ in the ordered and disordered phases as functions of $\rho_F$. It is clear that $\Delta$, the ordered-phase free energy, is positive for $0<\rho_F<0.3$, indicating that the disordered phase has a lower energy than the ordered phase, and so the Bose sector does not show superfluidity for $0<\rho_F<0.3$. In the Bose-dominated regime (small $\rho_F$), the system therefore prefers the phase with $\psi=0$. This is unlike the bosonic Hubbard model \cite{Sheshadri, PRL95, FisherWeichman89, bhmqmc}, where the system is a superfluid for $0<\rho_B<1$ in the hardcore ($U \rightarrow \infty$) limit. 

This behavior can be understood based on model (\ref{eq:ham}). In our approximation, the ground state energy per site of the model is $F_0 = -\mu(\rho_F) \rho_F - \phi \psi^2$. The filling constraint $\rho_F + \rho_B =1$ removes the $\alpha$-term and makes the chemical potential filling dependent, i.e. $\mu = \mu(\rho_F)$. The function $\mu(\rho_F)$ is plotted in Fig.\ref{fig:fig1}(b). It can be seen that $\mu(\rho_F)$ is negative for $\rho_F<1/2$, and thus the superfluid ground state energy can become positive in this regime. Numerically we find that $\Delta>0$ for $0<\rho_F<\rho_c \simeq 0.3$.

Simultaneously, for $\rho_F<\rho_c$, the Fermi sector shows an insulating phase, pointing to the important role of the coupling between Fermi and Bose sectors mediated by composite hopping. We note here that a zero-temperature insulating state with Cooper pairs has indeed been observed in amorphous Bismuth films \cite{valles2007}. 

\begin{figure}
\includegraphics[width=0.45 \textwidth]{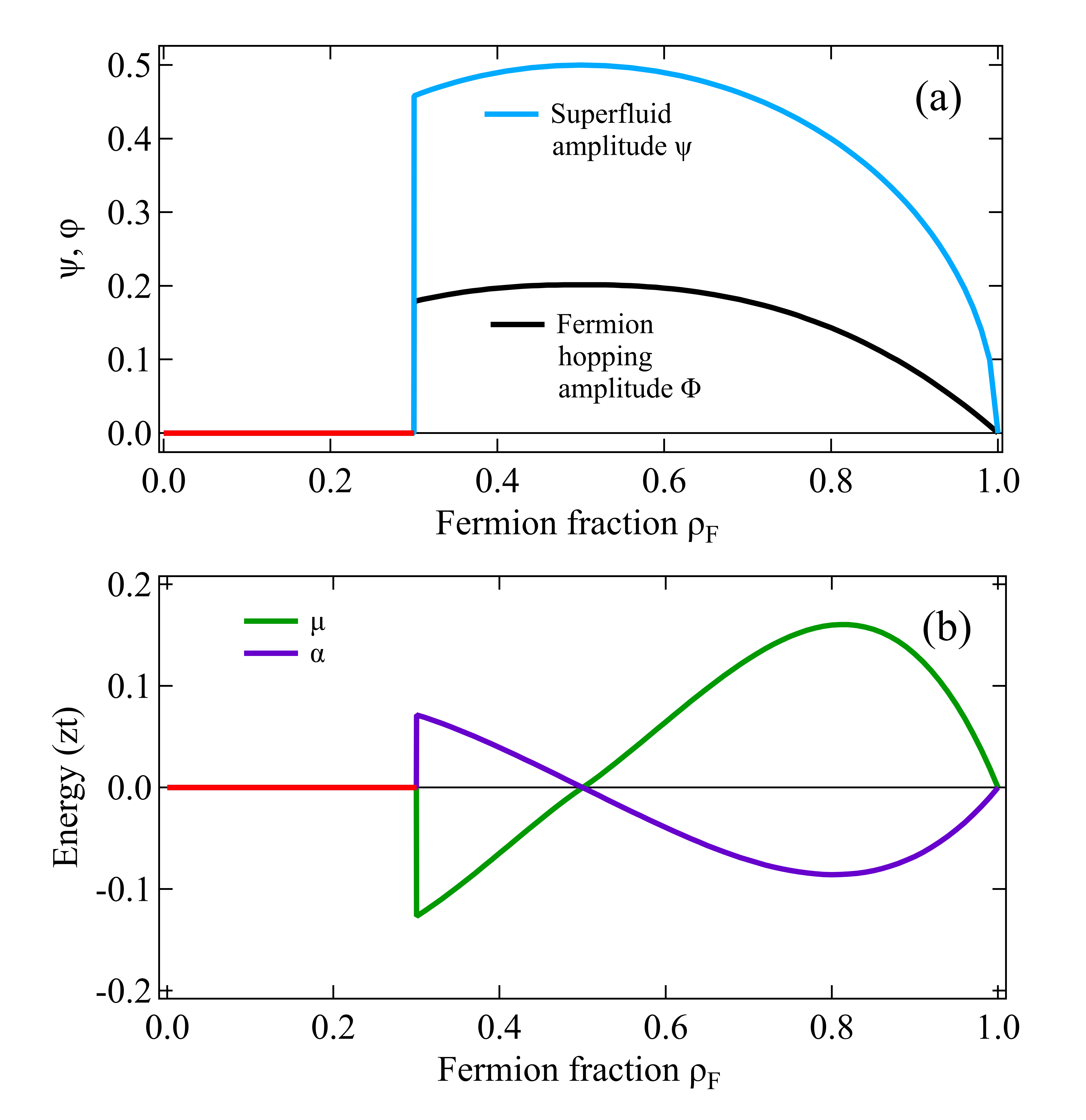}
\caption{Plots of (a) $\psi$ and $\phi$, and (b) $\mu$ and $\alpha$ as a function of $\rho_F$. At $\rho_F \simeq 0.3$, coupled first-order transitions are observed with jumps in  $\psi$, $\phi$, $\mu$ and $\alpha$.}	
\label{fig:fig3}
\end{figure}

\begin{figure}
\includegraphics[width=0.45 \textwidth]{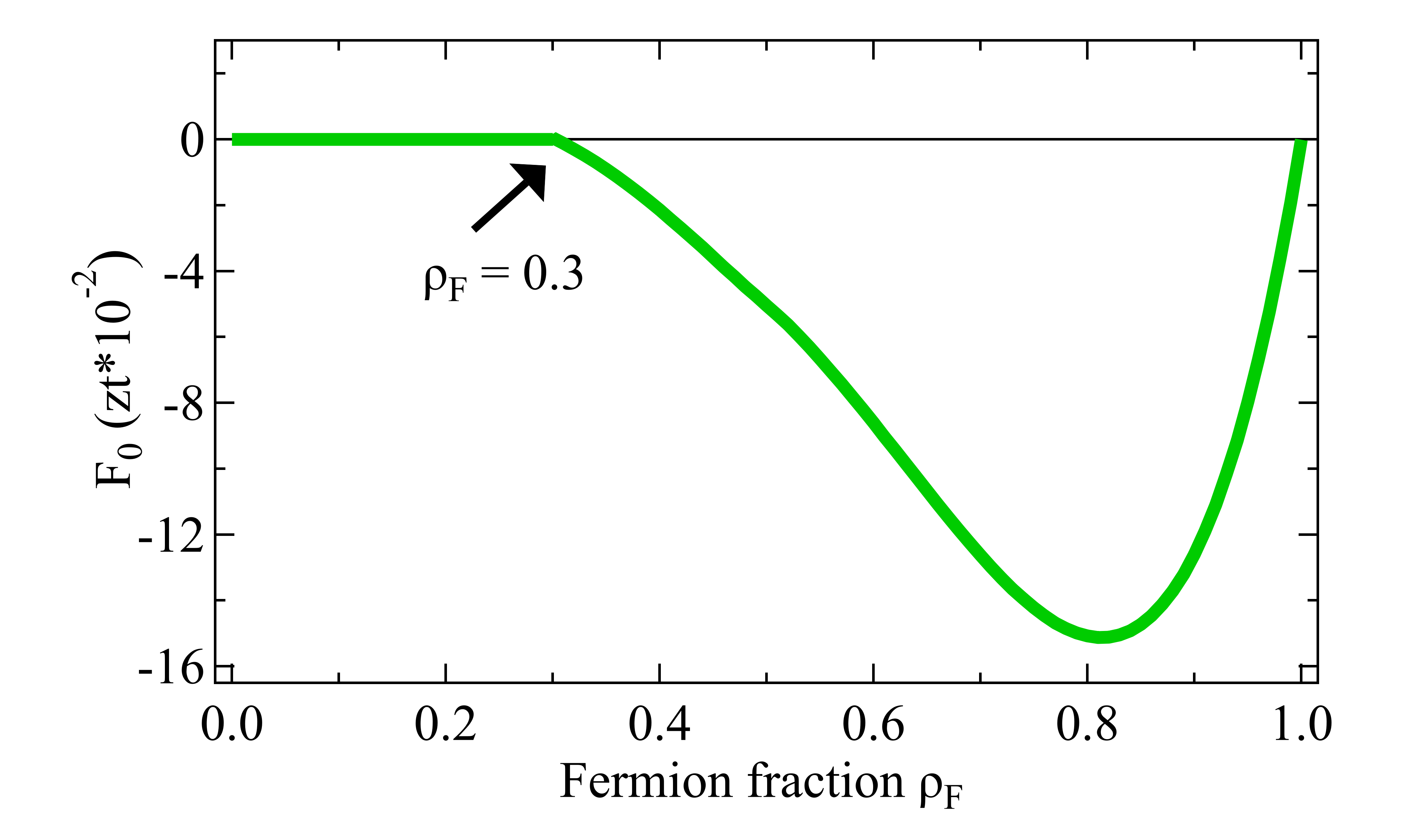}
\caption{Plot of the minimum free energy $F_0$, which is continuous with a derivative discontinuity at $\rho_F \simeq 0.3$.}	
\label{fig:fig4}
\end{figure}

From Fig.\ref{fig:fig2} it is clear that we must pick the disordered solution for $0<\rho_F<0.3$ and ordered solution for $0.3<\rho_F<1$ for each of the quantities shown in Fig.\ref{fig:fig1}. When we do this, we get results with discontinuities in $\psi, \phi, \alpha, \mu$ at $\rho_F=0.3$ as displayed in the figures \ref{fig:fig3}(a) and \ref{fig:fig3}(b). The free energy $F_0$ is continuous with a derivative discontinuity as shown in Fig.\ref{fig:fig4}. Since there are jumps in both the Bose ($\psi$) and Fermi ($\phi$) sectors, we have a coupled first-order transition from an insulating normal Bose gas ($\psi, \phi = 0$) to a metallic superfluid ($\psi, \phi > 0$), as we increase $\rho_F$ through $0.3$.

\begin{figure}
\includegraphics[width=0.45 \textwidth]{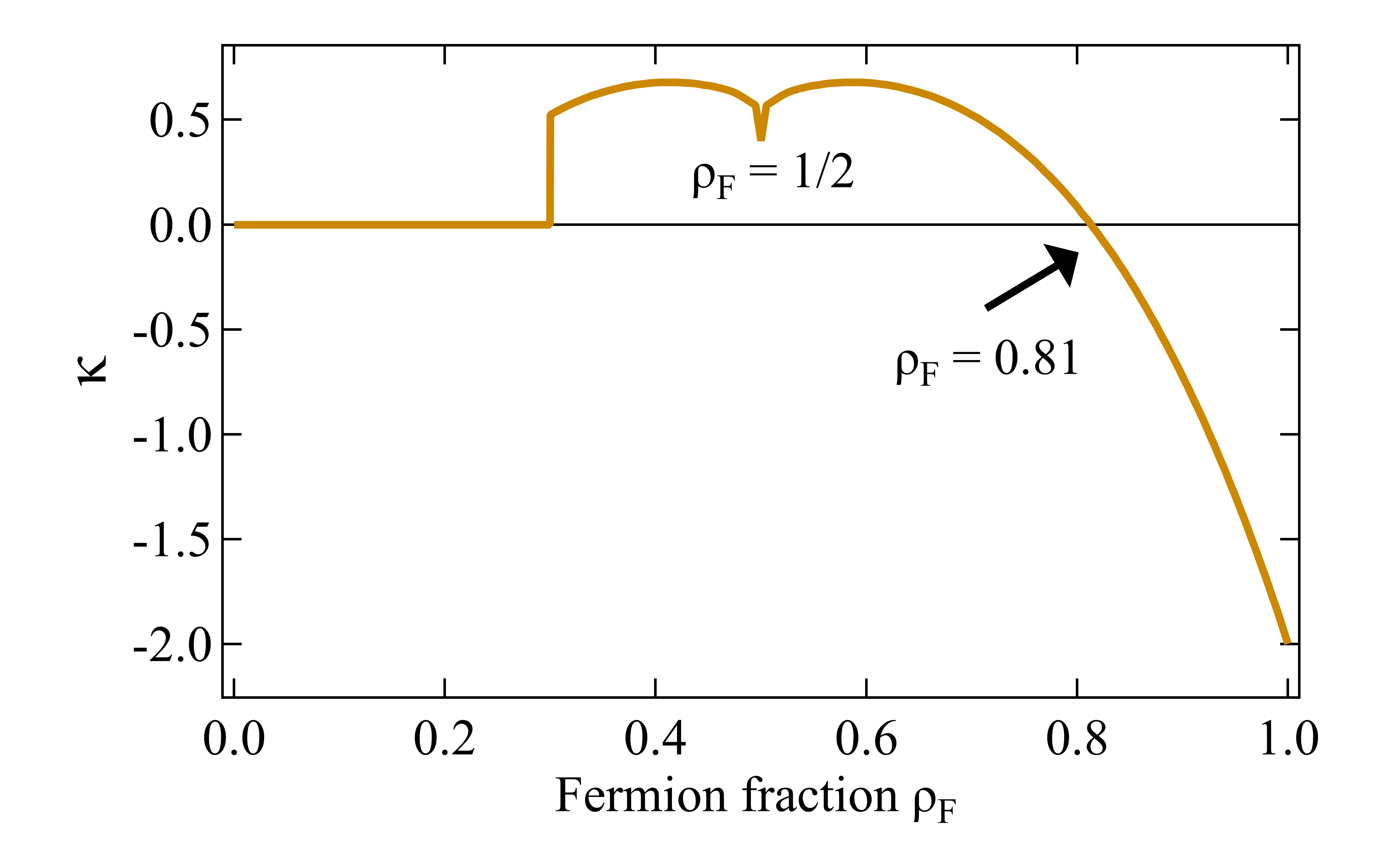}
\caption{Plot of the bulk modulus  $\kappa$ as a function of $\rho_F$. For $\rho_F>0.81$, $\mu$ and $F_0$ decrease with $\rho_F$, resulting in negative $\kappa$. The cusp at $\rho_F=1/2$ is a result of the van Hove singularity in the density of states.}
\label{fig:fig5}
\end{figure}

In Fig.\ref{fig:fig5}, we show a plot of the bulk modulus $\kappa = \partial \mu/\partial \rho_F = 2 [\phi + u_F (1-2 \rho_F)] + \pi^2\psi^2/[2 f(u_F)]$. The bulk modulus has a cusp-like behavior with a value of $2\phi$ at $\rho_F=1/2$. This is because of the van Hove singularity of the function $f(u)$ which is present in the last term of Bulk modulus above. It can also be seen that $\kappa$ vanishes for $\rho_F =\rho_{\kappa} \simeq 0.81$, becoming negative thereafter and approaching $-2$ at $\rho_F=1$. A negative value of $\kappa$ also corresponds to a negative compressibility of the fermions, which has been observed in two-dimensional electron gases \cite{negcomp_1} and materials with strong spin-orbit coupling \cite{He2015, Riley2015}. 

We have analyzed the factors responsible for the vanishing and negative values of bulk modulus, which is directly related to the function $\mu (\rho_F)$ becoming decreasing for $\rho_{\kappa}<\rho_F<1$. Firstly, $\mu = -\alpha + u_F \psi^2$ is determined by the filling constraint (\ref{eq:fill1}) that results in $\alpha (\rho_F) = (1-2\rho_F)\phi$. Secondly, we observe that the free energy minimum $F_0$ is zero at $\rho_F=0.3$ and $\rho_F=1$ and negative in between, and is therefore bound to have a minimum (at $\rho_F \simeq 0.81$). This minimum is the result of a competition between the Fermi sea and the composite hopping terms, as is clear from the expression $F_0 = -\mu(\rho_F) \rho_F - \phi \psi^2$ that we presented above. Since the composite hopping term ($- \phi \psi^2$) is decreasing for $\rho_F>1/2$, the $\mu$-term is forced to become decreasing to the right of the $F_0$ minimum, with a maximum at $\rho_F \simeq 0.81$.

\begin{figure}
\includegraphics[width=0.45 \textwidth]{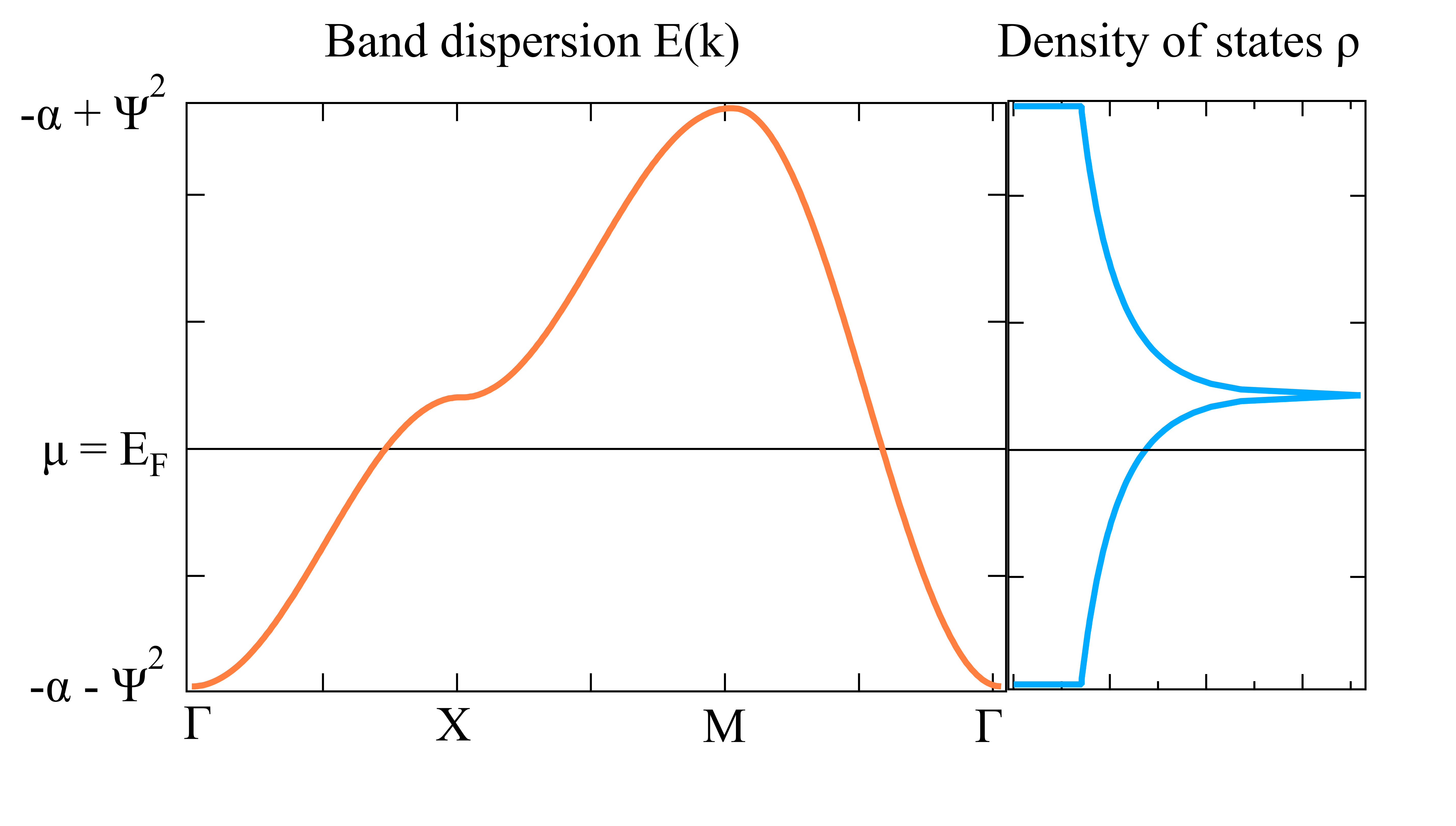}
\caption{Plots of (a) the Fermi band dispersion $\varepsilon_{\bf k}$ along $\Gamma$XM$\Gamma$ of the first Brillouin zone of the two-dimensional square lattice, and (b) the Fermi density of states $\rho(E)$, at an arbitrary value of $\rho_F$ in the metallic phase. The two plots share a common energy axis (ordinate).  Also marked is the Fermi energy $\mu$ at $\rho_F$ as a horizontal line between energies $-\alpha-\psi^2$ and $-\alpha+\psi^2$. It can be seen that the width of the Fermi band is $2\psi^2$, and therefore varies as the Fermi level (i.e. Fermi filling fraction) changes.}
\label{fig:fig6}
\end{figure}

Fig.\ref{fig:fig6} shows plots of Fermi band dispersion $\varepsilon_{\bf k}$ and density of states $\rho(E)$ at an arbitrary value of $\rho_F$. The Fermi band width is $2\psi^2$, completely determined by the superfluid density. Also marked is the location of the Fermi energy $\mu$ as a horizontal line. We can see that $\rho(E)$ has a cusp-like singularity corresponding to energy where $\nabla_{\bf k} \varepsilon_{\bf k}$ vanishes, and this is evidently a van Hove singularity. This singularity occurs at the Fermi energy at $\rho_F=1/2$, and this is where the superfluid amplitude has a maximum, as can be seen in Fig.\ref{fig:fig1}(a). The van Hove singularity is an important feature of the electronic structure of the high-T$_c$ cuprates \cite{vanhove_1, vanhove_2, Dessau, King, Gofron}. 

\begin{figure}
\includegraphics[width=0.45 \textwidth]{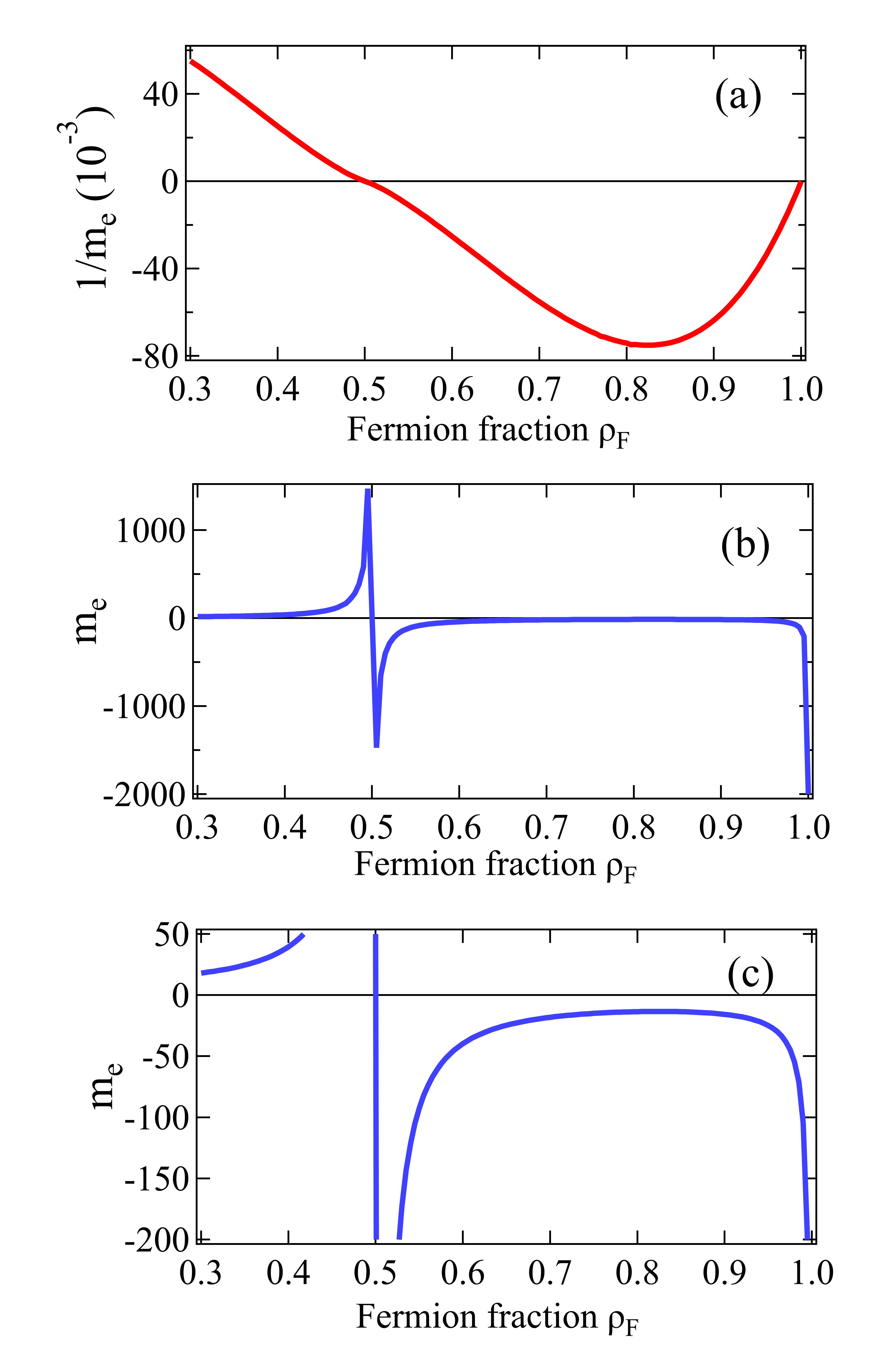}
\caption{Panel (a) shows the second-derivative of $\varepsilon_{\bf k}$ with respect to ${\bf k}$ (the reciprocal of fermion effective mass $m_{e}$) and panels (b) and (c) show plots of $m_{e}$ with two different vertical scales to highlight its behavior near and away from the singularities at $\rho_F$ = 1/2, 1, respectively. In (a), the minimum is at $\simeq 0.84$; this is slightly higher than where the maximum is ($\simeq 0.81$) in the chemical potential in Fig.1(b).}  	
\label{fig:fig7}
\end{figure}

Fig.\ref{fig:fig7}(a) shows a plot of the second-derivative of $\varepsilon_{\bf k}$ with respect to ${\bf k}$ at Fermi energy as a function of $\rho_F$ (see equation (9)), that becomes $-u_F\psi^2 \equiv -(\mu + \alpha)$ upon simplification. Figures \ref{fig:fig7}(b) and (c) show the behavior of the inverse of this second derivative, which is the Fermi effective mass $m_e$ on two different y-scales. The effective mass changes sign from particle like ($\rho_F<1/2$) to hole like ($\rho_F>1/2$), and has singularities at $\rho_F=1/2$ (the van Hove point where $u_F=0$) and $\rho_F=1$ (where $\psi=0$). More interestingly, the second-derivative of $\varepsilon_{\bf k}$ with respect to ${\bf k}$ shows a minimum at a certain $\rho_F$ slightly higher than $\rho_{\kappa}$, which also corresponds to a minimum of the hole effective mass. This is followed by a systematic increase in the effective mass for $\rho_F >\rho_{\kappa}$, culminating in a divergence at $\rho_F =1.0$.

\begin{figure}
\includegraphics[width=0.45 \textwidth]{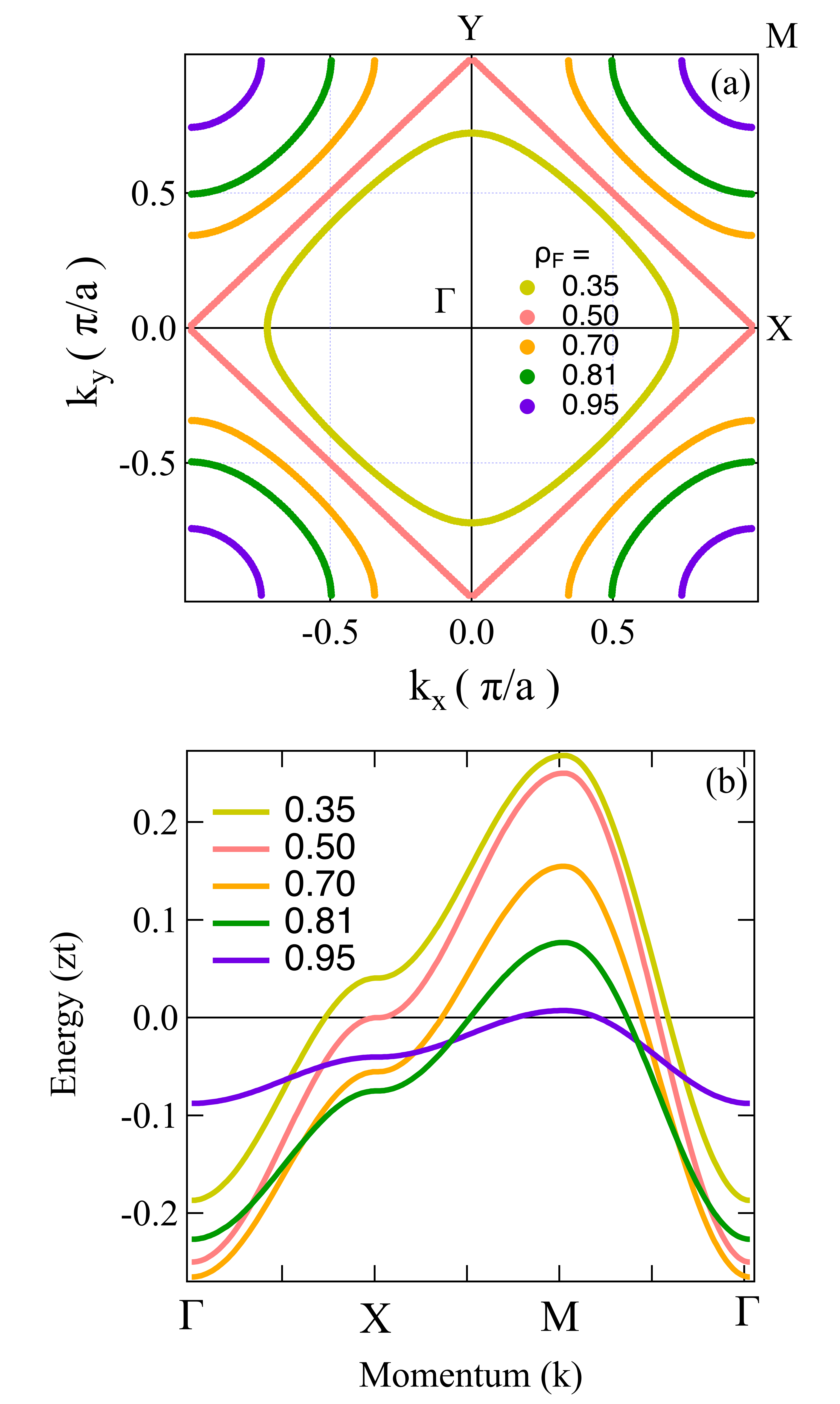}
\caption{(a) Plots of Fermi surface in the first Brillouin zone of the two-dimensional square lattice at five different values of $\rho_F$, namely $0.35, ~1/2, ~0.70, ~0.81, ~0.95$ in the metallic phase. It can be seen that the Fermi surface is convex (particle-like) for $\rho_F<1/2$, flat at $\rho_F=1/2$, and concave (hole-like) for $\rho_F>1/2$.
(b) Plot of band dispersions for $\rho_F$ = 0.35, 1/2, 0.7, 0.81 and 0.95, corresponding to the particle and hole Fermi surfaces shown in (a).}
 \label{fig:fig8}
 \end{figure}

The Fermi surface is determined by the equation $\varepsilon_{\bf k} = \mu$, that simplifies to $\gamma_{\bf k} = -u_F$. Fig.\ref{fig:fig8}(a) shows plots of Fermi surface in the first Brillouin zone at five different values of $\rho_F$, namely $0.35, ~1/2, ~0.70, ~0.81, ~0.95$, in the metallic phase. It can be seen that as a function of filling, the nature of carriers changes from particle-like for $\rho_F<1/2$ (convex surface at $\rho_F=0.35$) to hole-like for $\rho_F>1/2$ (concave surfaces at $\rho_F=0.70, ~0.81, ~0.95$), with a flat surface at $\rho_F=1/2$. This behavior is qualitatively similar to that of a tight-binding model. What is unusual, and distinguishes our model from the tight-binding model, is that the fermion bandwidth $2\psi^2$ changes as the chemical potential $\mu$ changes, as shown in the plots of figure \ref{fig:fig8}(b),
where we show the band dispersions for  $\rho_F$ = 0.35, 1/2, 0.7, 0.81 and 0.95. It is observed that the bandwidth is maximum at $\rho_F$ = 1/2, (which corresponds to the case of the van Hove singularity at the Fermi energy), and decreases for values of $\rho_F$ below and above that. It is also clear that the band dispersions  for  $\rho_F$ = 0.35 is particle-like and centered at the $\Gamma$-point, while for  $\rho_F$ = 0.7, 0.81 and 0.95, they are hole-like and centered at the M-point in the Brillouin zone.

\begin{figure}
\includegraphics[width=0.45 \textwidth]{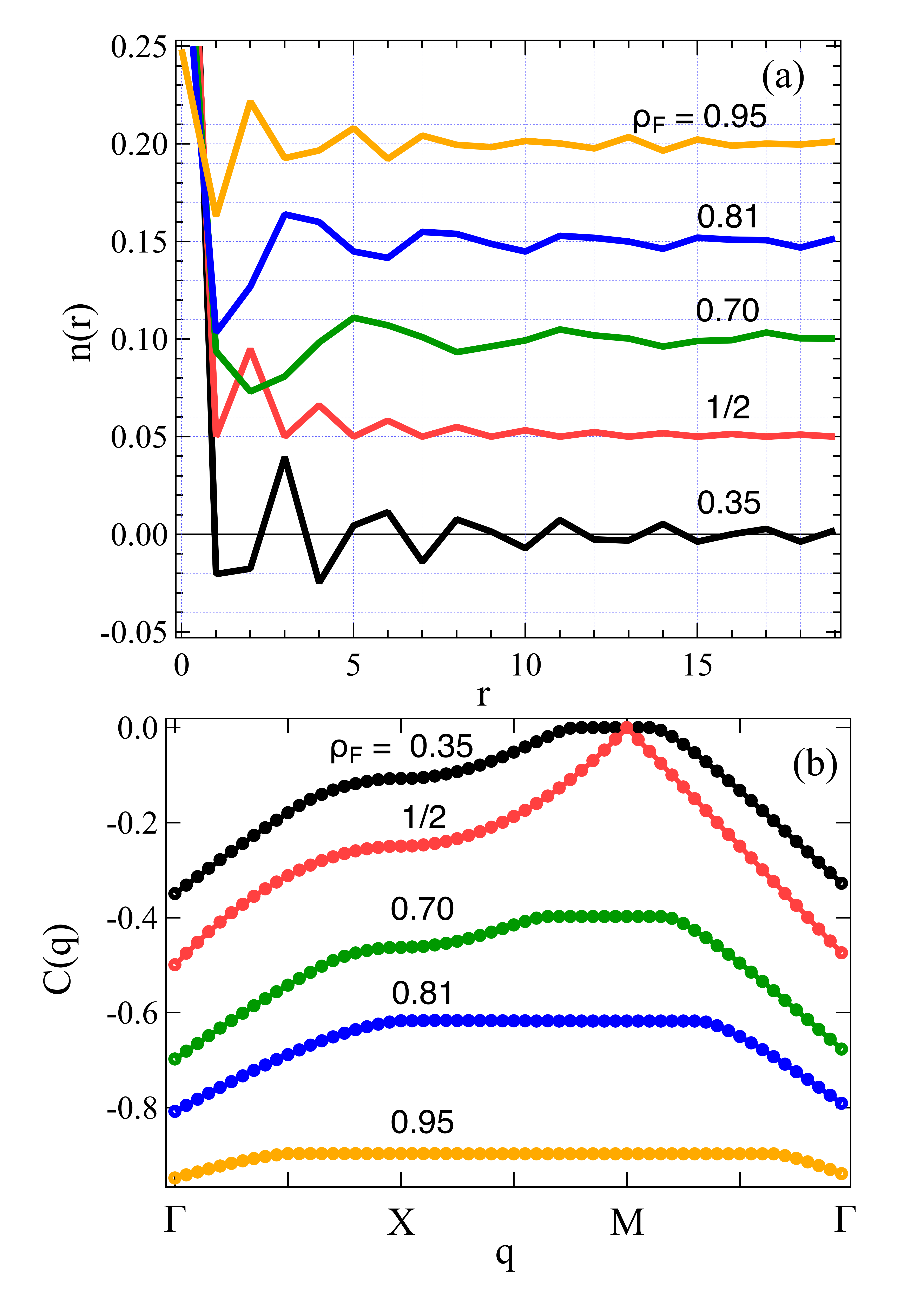}
\caption{(a) Plots of the function $n({\bf r})$ (see text, $C({\bf r}) = -n^2({\bf r})$) along the $x$-direction of the square lattice, and (b) of the Fourier transform $C({\bf q})$ along the                                $\Gamma$XM$\Gamma$ direction in ${\bf q}$-space, for $\rho_F = 0.35, ~1/2, ~0.70, ~0.81, ~0.95$ in the metallic phase. The curves in (a) are shifted along the y-axis for clarity. Observe the periodicity of $n({\bf r})$ for $\rho_F=1/2$, with a wavelength of twice the lattice spacing, indicating a density wave (DW) ordering in this case. The cusp in $C({\bf q})$ at the M-point for $\rho_F$ =~1/2 corresponds to the nesting vector along $\Gamma$-M direction }
 \label{fig:fig9}

\end{figure}

Finally, we consider the correlation function $C({\bf r}) = (1/N)\sum_i\langle \delta \rho_{{\bf r}_i} \delta \rho_{{\bf r}_i + {\bf r}} \rangle$ where $\delta \rho_{{\bf r}_i} = f_i^{\dagger}  f_i - \rho_F$ is the deviation from mean fermion density at lattice site ${\bf r}_i$. Using Wick's theorem we write the expectation value of a four-fermion operator as a sum of products of expectation values of two-fermion operators. We then obtain $C({\bf r}) = - n^2({\bf r})$ where $n({\bf r}) = (1/N) \sum_{\bf k} \langle  f_{\bf k}^{\dagger}  f_{\bf k}  \rangle  e^{i{\bf k}.{\bf r}}$, and $C_{\bf q} = -(1/N) \sum_{\bf k} \langle  f_{\bf k}^{\dagger}  f_{\bf k}  \rangle  \langle  f_{\bf k+q}^{\dagger}  f_{\bf k+q}  \rangle$ for the Fourier transform of $C({\bf r})$. Fig.\ref{fig:fig9} shows the plots of $n({\bf r})$ and $C_{\bf q}$ for several values of $\rho_F$. $n({\bf r})$ has been plotted along the $x$-axis as a function of lattice spacing. It has a periodicity of twice the lattice spacing for $\rho_F=1/2$, but no apparent periodicity for other values of $\rho_F$. This is also supported by the plot of  $C({\bf q})$ in (b) that shows a cusp-like singularity for $\rho_F=1/2$ at the M-point with ${\bf q}=(\pi, \pi)$, but is nearly featureless for other values of $\rho_F$.

For $\rho_F = 1/2$, we can analytically evaluate the ${\bf k}$-space integrals to obtain $\pi^2 n({\bf r}) = (\cos \pi y - \cos \pi x)/(x^2 - y^2)$. This shows that there is a density wave (DW) with wave vector ${\bf Q} = (\pi, \pi)$: the square lattice is divided into two sub-square lattices with twice the lattice constant, with fermions occupying one of them and bosons, the other. In this case, every fermion has a boson as its nearest neighbor and vice versa. This enhances composite hopping and leads to maximum superfluidity and metallicity, as we saw in Fig.\ref{fig:fig1}(a).

We write $\langle  f_{\bf k}^{\dagger}  f_{\bf k}  \rangle \equiv g(\varepsilon_{\bf k})$, where the function $g(x)$ is 1 if $x \le \mu$ and 0 otherwise, so that $C_{\bf q} = -(1/N) \sum_{\bf k} g(\varepsilon_{\bf k}) g(\varepsilon_{\bf k+q})$. For $\rho_F=1/2$, we have $\alpha=0$, so from equation (\ref{eq:dispersion}) we have $\varepsilon_{\bf k+Q} = -\varepsilon_{\bf k}$. For ${\bf q=Q}$, we can use this nesting property of the Fermi surface and convert the ${\bf k}$-sum to an integral over the density of states to obtain $C_{\bf Q} = -\rho (0) (\mu-0) = 0$, in agreement with Fig.\ref{fig:fig9}(b). This points to the role of Fermi surface nesting in causing a DW order for $\rho_F=1/2$.

The filling $\rho_F=1/2$ turns out to be very special: we have a van Hove singularity resulting in a cusp in $\kappa$, and Fermi surface nesting resulting in a DW phase coexisting with a maximum in Bose superfluid and Fermi hopping amplitudes. In this case, $F_0 \simeq 0.05 ~zt$ can be taken to be a reasonable approximation of $T_c$, the critical temperature for BEC. To estimate this, we consider a situation with two neighboring sites on a lattice, one with a single electron with spin orientation $S_z = +1/2$ or $-1/2$, and the other with a pair of electrons, one with $S_z=1/2$ and the other, $S_z=-1/2$; such a pair can be treated as a boson \cite{Schafroth, Maska2}. The hopping of one of the electrons of the pair to the singly-occupied site in this situation is the same as composite hopping in model (\ref{eq:ham}). This could be realized in a strongly-correlated material with a narrow single-band above half filling, in which hopping strength can be typically $\sim 0.5$eV, and that would suggest a BEC transition temperature of several hundred kelvins in such a case. For simplicity we have ignored fermion spin in our model, and plan to include it in the near future. For ultracold atom systems, however, the hopping amplitude might be in the subkelvin range, and we can expect $T_c$ to be of the same order.

We note a few unusual features of the insulating and metallic phases of the Fermi sector. We have insulating behavior at a range of {\em partial} fillings of the Fermi band. This is unlike the usual case where the insulating behavior is associated with a band gap. In our model, the insulating behavior of the Fermi sector results from the energy cost of superfluid phase of the Bose sector. This results in the insulating Fermi sector for $0<\rho_F<0.3$. In contrast, the insulating behavior for $\rho_F=1$ results from a divergent effective mass $m_e$ due to narrowing of the fermion band (see Fig.\ref{fig:fig7}). The metallic phase close to the van Hove point $\rho_F=1/2$ is also unusual because the metallicity (as measured by $\phi$) is maximum while $m_e$ has a large magnitude.

\section{Conclusion}
\label{conclude}

In conclusion, we have presented a model of FBMs with a hopping mechanism that exhibits several interesting properties at zero temperature. In the Bose-dominated regime, the model is in an insulating normal phase: the Fermi sector is insulating, while the Bose sector is a normal gas. As the Fermi filling fraction is increased, the model has coupled first-order transitions at $\rho_F \simeq 0.3$, where superfluid amplitude $\psi$ and fermion hopping amplitude $\phi$ jump to finite nonzero values. For $\rho_F>0.3$, the system is a metallic superfluid: the Fermi sector is metallic, while the Bose sector is a superfluid, with maximum values of $\psi, ~\phi$ coexisting with a DW order at $\rho_F = 1/2$. There is a van Hove singularity in the density of states at Fermi energy that makes the bulk modulus $\kappa$ develop a cusp-like minimum for $\rho_F = 1/2$; $\kappa$ vanishes at $\rho_F \simeq 0.81$, and is negative for $\rho_F>0.81$, as a result of a competition between Fermi sea and  composite-hopping contributions to the ground-state energy. The fermion band width varies with the chemical potential with a maximum at $\rho_F = 1/2$, the van Hove point. The fermion effective mass displays singular behavior at $\rho_F =1/2, ~1$. And finally, our estimate for the BEC transition temperature is in the subkelvin range for ultracold atom systems and several hundred kelvins in possible solid-state instances of composite hopping. 


\end{document}